\documentclass[aps,pre,preprint,groupedaddress,
,superscriptaddress,showpacs,a4paper,floatfix]{revtex4-1}

\usepackage{amsmath,amssymb,graphicx,graphics,bm}
\usepackage{hyperref}
\usepackage{graphics,graphicx}
\usepackage{amsmath,amssymb}
\usepackage{color}

\begin{document}
\title{Chemical kinetics and stochastic differential equations}

\author{Chiara Pezzotti and Massimiliano Giona}
\email[corresponding author:]{massimiliano.giona@uniroma1.it}
\affiliation{Dipartimento di Ingegneria Chimica, Materiali, Ambiente La Sapienza Universit\`a di Roma\\ Via Eudossiana 18, 00184 Roma, Italy}

\date{\today}

\begin{abstract}
We propose a general stochastic formalism for describing the
evolution of chemical reactions  involving a finite number of molecules.
This approach is consistent with the statistical analysis based on
the Chemical Master Equation, and provides the formal setting for 
the  existing algorithmic approaches (Gillespie algorithm).
Some practical advantages of this formulation are  addressed, and
several examples are discussed pointing out the
 connection with quantum transitions (radiative interactions).
\end{abstract}

\maketitle

All the chemical physical processes involve, in  an atomistic perspective, 
a stochastic description of the events,  be them reactive or associated with
a change of phase (for instance adsorption) \cite{chemfluct}.
Nonetheless, in the overwhelming majority of the cases of practical and laboratory
interest, the number of molecules involved  is so large  to justify 
a mean field approach,
essentially based on the Boltzmannian hypothesis of molecular chaos (the
``stosszahlansatz'') \cite{boltzmann}.
The mean field formulation represents the backbone of  the classical theory
of chemical reaction kinetics \cite{chemkin1,chemkin2}.

It is well known that, in all the cases  where the number of molecule
is small (and this occurs in subcellular biochemical reactions,
in nanoscale systems, or in the growth kinetics of microorganisms \cite{bio1, bio2, bio3}),
the effects of fluctuations become significant, motivating a stochastic
description of chemical kinetic  processes,
involving the  number of molecules present in the system, thus
explicitly accounting for  due to their finite number  \cite{mcquarrie,stoca,
stoca1,stoca2}.
The statistical theory of chemical kinetics in these conditions
is grounded on the Chemical Master Equation (CME) \cite{gillespierigorous,cme3},
expressing the evolution equation for the probabilities $p({\bf N},t)$
of all the possible number-configurations ${\bf N}(t)=(N_1(t),\dots,N_s(t))$,
where $N_h(t)$ is the number of molecules of the $h$-th reacting species
at time $t$, $h=1,\dots,s$.
However, apart from a handful of simple cases, for which the CME can be
solved analytically \cite{cmecpl}, numerical methods should be applied to it in order to compute mean values and higher-order moments. But also
this choice reveals itself  to be unfeasible in most of the situations 
of practical
and theoretical interests, due to the extremely large number of configurations involved, making the multi-index matrix $p({\bf N},t)$ so huge to
exceed  reasonable computational facilities.

In order to solve this problem, Gillespie proposed an algorithmic solution
to the numerical simulation  of stochastic reacting systems, based
on the Markovian nature of the reactive events \cite{gillespiegeneral,gillespieexamples}. The original Gillespie algorithm
has been extended and improved over time, providing a variety of
slightly different computational alternatives.
A common denominator of the first family of the Gillespie algorithms 
(namely those based on the direct method, the first reaction method or their 
derivates \cite{ff1,ff2,ff3}) is to associate to every time step the 
occurrence of just one reaction. This formulation comes directly from the 
assumption that, if the time step is small enough, the probability that more 
than one reaction will occur is negligible. While correct, this choice brings 
to significant computational costs for complex reaction schemes. 
This problem has been highlighted several times, from the Gillespie group 
itself, as \emph{stiffness} in stochastic chemical reacting systems 
\cite{stiffness}. A brilliant way to overcome this limit  originates 
the famous tau-leaping method, which, unfortunately, requires to 
check that the propensity functions remain \emph{almost constant} 
at each iteration and can be applied just if this condition
is verified \cite{tau1,tau2}. The algorithmic solution associated with the formalism here introduced combines the accuracy of the first SSA with the computational advantages of the $\tau$-leaping method.

There is, moreover, a missing link between the CME theory and the Gillespie
algorithm, consisting in the straight mathematical
formulation of the stochastic differential equations associated with
a chemical reacting system, the statistical description of which would
correspond to the CME.
To clarify this issue, consider the  conceptually analogous problem
of  particle diffusion over the real line, the statistical
description of which is expressed by the parabolic equation $\partial p(x,t)/\partial t=D \, \partial^2 p(x,t)/\partial x^2$, for the probability
density  $p(x,t)$ of finding a particle at position $x$  at time $t$.
Setting $x_n=x(n \, \Delta t)$, an algorithm describing this process can
be simply expressed by the discrete evolution equation $x_{n+1}=
x_n + \sqrt{2 \, D \, \Delta t} \, r_{n+1}$, where $r_h$, $h=1,2,\dots$
represent independent random variables sampled from
a normal distribution (with zero mean, and unit variance) \cite{diffusion}. This
represents an efficient algorithmic solution of the problem,  whenever 
the time resolution $\Delta t$  is small enough. Nevertheless, the mere algorithmic approach cannot be considered physically satisfactory, in 
a comprehensive formulation of transport theory
embedded in a continuous space-time (in which both position $x$
and time $t$ are real valued). In point of fact, only with the  mathematical
formulation
due to K. Ito of stochastic differential equations driven by the
 increments $dw(t)$ of a Wiener process (Langevin equations) \cite{ito}, 
namely $d x(t)= \sqrt{2 \, D} \, d w(t)$  the theory of diffusive 
motion has found a proper mathematical physical setting.

A similar situation applies to the case of stochastic models
of chemical reaction kinetics, and the present Letter is aimed at filling this
gap.
The basic idea is that any reactive process corresponds to a system of 
elementary events (the single reaction) possessing a Markovian transitional
structure, 
and, consequently, amenable to a description by means of the increments of
counting processes (Poisson processes, in the Markovian case).
This topic has been also pointed out in \cite{closed1} in terms
of Poisson measures,  although the latter formulation is much less
simple and physically intuitive  than the approach  proposed in the present Letter.

\begin{figure}
\includegraphics[width=15cm]{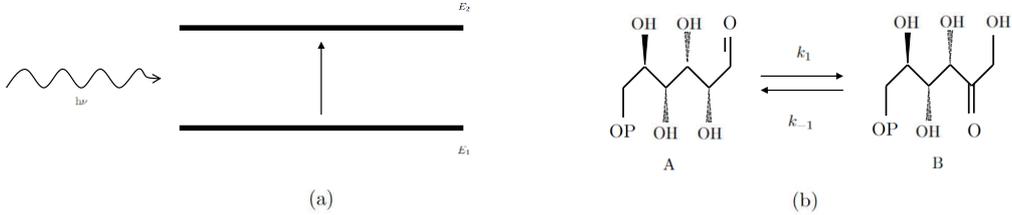}
\caption{Schematic  representation of the analogy between a 
two-level quantum system
and a first-order chemical kinetics, such as an isomerization.}
\label{Fig0}
\end{figure}

To begin with, consider the simple case of a first-order chemical
reaction $A \underset{k_{-1}}{\stackrel{k_1}{\rightleftharpoons}} B$ (for instance,
an isomerization). This model is perfectly analogous to the radiative 
transition
of a molecule possessing two energy states, due to emission and adsorption of an energy quantum (figure \ref{Fig0}). Let $N_A(0)+N_B(0)=N_g$ the total number of molecules at time $t=0$. The state of the system is characterized by the
state functions $\sigma_h(t)$, $h=1,\dots,N_g$ for each molecule, attaining values $\{0,1\}$, and such that $\sigma_h(t)=0$ if the energy state at time $t$ is $E_0$ (or equivalently if the molecule finds itself in the state $A$), and $\sigma_h(t)=1$ in the opposite case (energy state $E_1$, or isomeric state $B$).

Let $\{ \chi_h^{(1)}(t,k_1), \chi_h^{(2)}(t,k_{-1}) \}_{h=1}^{N_g}$ be two
systems of independent Poisson processes, characterized by the transition rates $k_1$, and $k_{-1}$, respectively. The evolution of $\sigma_h(t)$ can be expressed via the stochastic differential equation
\begin{equation}
\frac{d \sigma_h(t)}{d t}= \left (1- \sigma_h(t) \right ) \, \frac{d \chi_h^{(1)}(t,k_1)}{d t} -  \sigma_h(t) \,  \frac{d \chi_h^{(2)}(t,k_{-1})}{d t}
\label{eq1}
\end{equation}
$h=1,\dots,N_g$, where $d \chi(t,\lambda)/d t$ is the distributional
derivative of the Poisson process $\chi(t,\lambda)$, corresponding to a sequence of
unit impulsive functions at the transition instants $t_i^*$, $i=1,2,\dots$,
$0<t_{i}^* < t_{i+1}^*$, where for  $\varepsilon>0$, $\lim_{\varepsilon  \rightarrow 0}  \int_{t_i^*-\varepsilon}^{t_i^*+\varepsilon} d \chi(t,\lambda)=1$. Summing over $h=1,\dots N_g$, and observing that $N_A(t)= \sum_{h=1}^{N_g} \left (1- \sigma_h(t) \right )$,
$N_B(t)= \sum_{h=1}^{N_g} \sigma_h(t)$, we have
\begin{equation}
\frac{d N_B(t)}{ dt} = \sum_{h=1}^{N_A(t)} \frac{d \chi_h^{(1)}(t,k_1)}{d t}
- \sum_{h=1}^{N_B(t)} \frac{d \chi_h^{(2)}(t,k_{-1})}{d t}
\label{eq2}
\end{equation}
and $d N_A(t)/d t=- d N_B(t)/d t$, representing the evolution equation
for $N_A(t)$ and $N_B(t)$, attaining integer values. The stochastic evolution
of the number of molecules $N_A(t)$, $N_B(t)$ is thus
 expressed as a differential equation with respect to the continuous physical time $t \in {\mathbb R}^+$,
over the increments of a Poisson process.
Intepreted in a mean-field  way, if $c_{\rm tot}$ is the overall concentration
of the reactants at time $t=0$, then the concentrations
$c_\alpha(t)$ at time $t$ can be recovered from eq. (\ref{eq2}) as 
\begin{equation}
c_\alpha(t)= c_{\rm tot} \, \frac{N_\alpha(t)}{N_g} \, , \qquad \alpha=A,B
\label{eq3}
\end{equation}
representing the calibration relation connecting the stochastic
description in terms of number of molecules $N_\alpha(t)$ and  the
concentrations $c_\alpha(t)$, $\alpha=A,\,B$ entering the mean-field description.

The analytical formulation of a stochastic differential equation for chemical 
kinetics, expressed in terms of the number of molecules of the chemical
species involved, rather than an algorithm  defined for discretized times,
permits to develop a variety of different numerical strategies, that naturally
 perform a modified tau-leaping procedure, as the occurrence of several distinct reactive events  in any elementary time step $\Delta t$ is intrinsically accounted
for.
This can be easily seen by considering the simple reaction defined
by the evolution equation (\ref{eq2}).
In terms of increments, eq. (\ref{eq2}) can be written as
$d N_B(t)= \sum_{h=1}^{N_A(t)} d \chi^{(1)}(t,k_1) - \sum_{h=1}^{N_B(t)}
d \chi^{(2)}(t,k_{-1})$. If $\Delta t$ is the chosen time step,
it follows  from this formulation, a  simple  numerical approximation for eq. (\ref{eq2}), namely,
\begin{equation}
\Delta N_B(t)=N_B(t+\Delta t)-N_B(t)=
\sum_{h=1}^{N_A(t)} \xi_h^{(1)}(k_1\, \Delta t) - \sum_{h=1}^{N_B(t)}
\xi_h^{(2)}(k_{-1} \, \Delta t)
\label{eq_xadd1}
\end{equation}
where $\xi^{(1)}(k_1 \, \Delta t)$, $\xi_h^{(2)}(k_{-1} \, \Delta t)$ $h=1,2,\dots$, are two families of independent binary random variables, where
\begin{equation}
\xi^{(\alpha)}_h(p) =
\left \{
\begin{array}{lll}
1 \;\; \; & \;\;& \mbox{with probability} \; p \\
0 & & \mbox{otherwise}
\end{array}
\right .
\label{eq_xadd2}
\end{equation}
$\alpha=1,2$, $h=1,2,	\dots$.
The time step $\Delta t$, can be chosen in eq. (\ref{eq_xadd1}) from the
condition
\begin{equation}
K \, \Delta t < 1 \, , \qquad K=\max \{k_1,k_{-1} \}
\label{eq_xadd3}
\end{equation}
In practice, we choose $\Delta t=0.1/K$. As can be observed, the choice of
$\Delta t$ is limited by the intrinsic rates of the process.
The advantage of deriving different algorithmic schemes for solving 
numerically  the stochastic kinetic  equations becomes more evident
in dealing with bimolecular reactions (addressed below). Due to
the intrinsic limitations of this communication, a  thorouh discussion
of this issue is postponed  to a future more extensive article \cite{PG_next}.

\begin{figure}
\includegraphics[width=10cm]{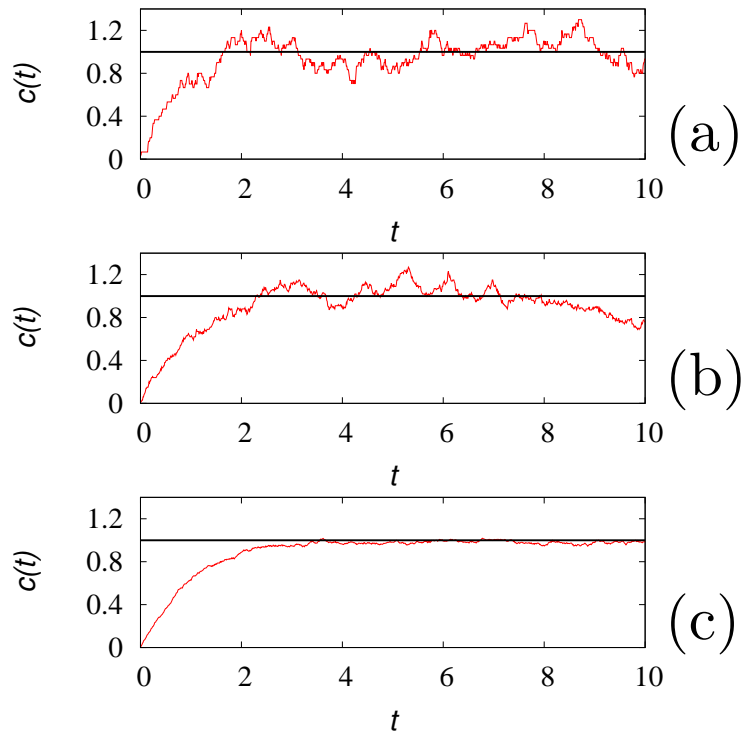}
\caption{$c(t)=N(t)/N_g$  vs $t$ from a single realization of the tank-loading process
eq. (\ref{eq5}) with $D=1$, $c_{0}=1$ a.u.. Panel (a): $N_g=30$, panel (b) $N_g=100$, panel (c) $N_g=1000$.
The solid horizontal lines represent the steady-state value $c^*=1$.}
\label{Fig1}
\end{figure}

\begin{figure}
\includegraphics[width=10cm]{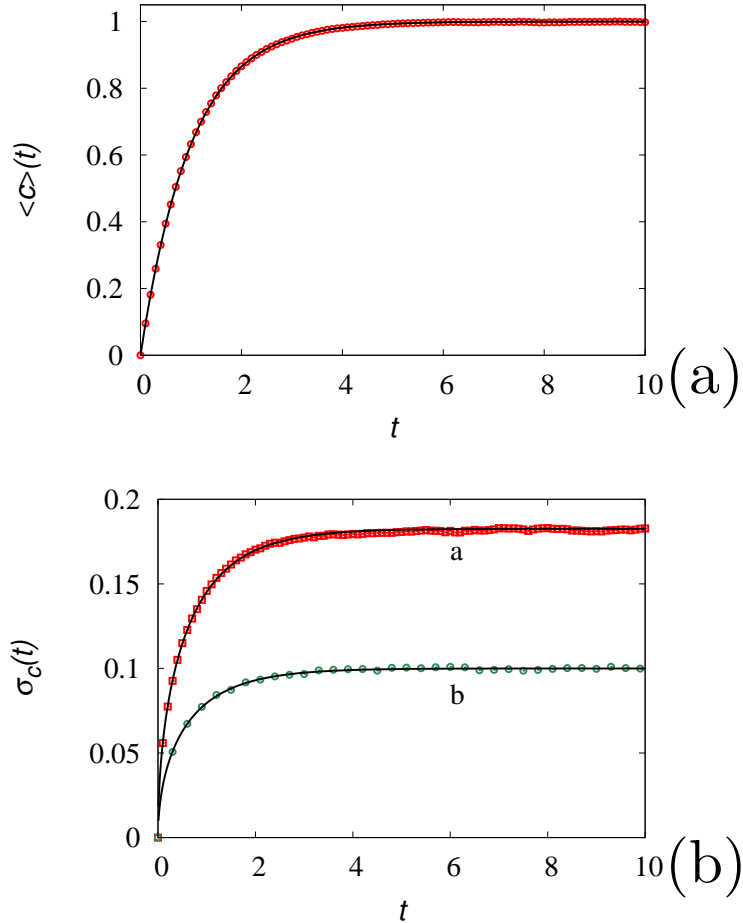}
\caption{Panel (a): $\langle c \rangle(t)$ vs $t$ at $N_g=30$ (symbols) averaged over $[ 10^6/N_g]$ realizations
of the tank-loading process with $D=1$, $c_0=1$ a.u. Here, $[\cdot]$ indicates the integer part of its argument. The solid line represents the mean-field 
result $\langle c \rangle(t)=1-e^{-t}$.
Panel (b): Variance  $\sigma_c(t)$ vs $t$ for the tank-loading process. Symbols are
the results of stochastic simulations of eq. (\ref{eq5}) averaged over $[ 10^6/N_g]$ realizations,
 lines the solutions
of eq. (\ref{eq7}). Line (a) refers to $N_g=30$, line (b) to $N_g=100$.}
\label{Fig2}
\end{figure}

The same approach can be extended to include amongst the elementary events
not only the reactive steps, but also feeding conditions, thus representing the evolution of chemically reacting systems with a finite number of molecules in a perfectly stirred open reactor. This is
the case of the tank-loading problem, in which a tracer 
is injected 
in an open vessel 
assumed perfectly mixed, for which, in the absence of chemical reactions,
the mean field equation for the concentration of the tracer reads
\begin{equation}
\frac{d c(t)}{d t}= D \, \left ( c_0 - c(t) \right )
\label{eq4}
\end{equation}
where $c_0$ is the inlet concentration and $D$  the dilution rate
(reciprocal of the mean retention time), and $c(0)=0$. 
Fixing $N_g$ so that $c(t)= c_0 \, N(t)/N_g$, the corresponding stochastic differential equation for the integer $N(t)$ involves, also in this case, two  families of counting processes, one  for the loading at constant concentration 
$c_0$, and 
the other for tracer discharge in the outlet stream, 
characterized by the same transition
rate $D$,
\begin{equation}
\frac{d N(t)}{d t}= \sum_{h=1}^{N_g}  \frac{d \chi_h^{(1)}(t,D)}{d t}
- \sum_{k=1}^{N(t)} \frac{d \chi_h^{(2)}(t,D)}{d t}
\label{eq5}
\end{equation}
starting from $N(0)=0$.
Figure \ref{Fig1} depicts several realizations of the tank-loading
process, obtained by discretizing eq. (\ref{eq5}) with a time step $\Delta t=10^{-3}$.
Despite the simplicity of the  process, this example permits to
highlight the role of $N_g$, that can be referred to as the {\em granularity number}, and the way stochastic models of chemical reactions can be fruitfully applied. Indeed, there is a two-fold use of the stochastic formulation of
chemical kinetic schemes. The first refers to a chemical reacting system
involving a small number of molecules, and in this case $N_g$ represents the effective number of molecules present in the system.
The other is  to use stochastic algorithms for simulating reacting systems 
in an alternative (and  sometimes more
efficient way) with respect to the solution of the corresponding mean-field equations. In the latter case, the granularity number $N_g$ represents essentially
a computational parameter, tuning the intensity of the fluctuations.
Two choices are then possible: (i) it can be chosen large enough, in 
order to obtain from a single realization of the process an accurate
approximation of the mean-field behavior, or (ii) it can be chosen small enough 
in order, to deal with extremely fast simulations of  a single realization of 
the process, that could be  averaged over a statistically significant
number of realizations in due time. These two choices are depicted in
figure \ref{Fig1} (panel c), choosing $N_g=10^3$, and in figure \ref{Fig2}
panel (a) obtained for $N_g=30$.
Of course, the latter approach is valid as long as the low-granularity
(low values of $N_g$) does not influence the qualitative properties of the
kinetics.

The second (computational) use  of stochastic simulations of chemical
kinetics requires a further discussion. At a first sight, it may appear that any stochastic simulation would be computationally less efficient than the solution of the corresponding mean-field equations. This is certainly true
for classical chemical reaction schemes in a perfectly mixed 
system, for which the mean-field model reduces to a system of ordinary
differential equations for the concentrations of the reactants. But there are
kinetic problems e.g., associated with the growth of microorganisms and eukaryotic cell lines in bioreactors 
(these growth phenomena, are indeed amenable to  a description in terms of equivalent chemical reactions),
the mean-field model of which is expressed in the form of  higher-dimensional
nonlinear
integro-differential equations . For this class of problems, 
the use of stochastic simulations
is the most efficient, if not the only way to achieve a quantitative
description of the process, in those cases where the number  $n_p$ of internal parameters describing the physiological state of an eukaryotic cell becomes large enough, 
$n_p \geq 3$. This issue is addressed in detail in \cite{PBGbio}.
This case is altogether similar to some transport problems, such as Taylor-Aris
dispersion for high P\'eclet numbers or  the analysis of 
 microfluidic separation processes (DLD devices) for which the stochastic simulation of particle motion  is far more efficient that the corresponding solution
of the  corresponding mean-field model expressed in the form
of advection-diffusion equations \cite{taris,dld}.

To complete the analysis  of the tank-loading problem, the associated CME reads
\begin{equation}
\frac{d p(n,t)}{d t}= D \, N_g \, \left [ p(n-1,t) \, \eta_{n-1} - p(n,t) \right ] + D \left [ (n+1) \, p(n+1,t) - n \, p(n,t) \right ]
\label{eq6}
\end{equation}
where $\eta_{h}=1$ for $h \geq 0$ and $\eta_h=0$ otherwise.
It follows  that $\langle c\rangle(t)=c_0 \sum_{n=1}^{\infty} n \, p(n,t)/N_g$
satisfies identically the mean-field equation (due to the linearity of the problem), while the variance $\sigma_c(t)$, with $\sigma_c^2(t)= c_0^2 \sum_{n=1}^\infty n^2 \, p(n,t)/N_g^2 - \left ( c_0 \sum_{n=1}^{\infty} n \, p(n,t)/N_g
\right )^2$, satisfies the equation
\begin{equation}
\frac{d \sigma_c^2}{d t}= - 2 \, D \, \sigma_c^2 + D \, \left ( \frac{1}{N_g} + \frac{\langle c \rangle}{N_g} \right )
\label{eq7}
\end{equation}
Figure \ref{Fig2} panel (b) compares the results of stochastic simulations
against the solutions of eq. (\ref{eq7}) for two values of $N_g$.

The above approach can be extended to any system of nonlinear reaction schemes involving unimolecular and bimolecular reaction, and  in the presence of slow/fast
kinetics. 
The structure of the reaction mechanism can be arbitrarily complicated without adding any  further complexity (other than purely notational)
in the formulation of the   stochastic evolution  expressed in terms of number of molecules.
The only practical issue, is that the number of different families of stochastic processes grows with
the number of elementary reactive processes considered.
For instance, in the case of  the subtrate-inhibited Michaelin-Menten kinetics
\begin{eqnarray}
E  +   S \underset{k_{-1}}{\stackrel{k_1}{\rightleftharpoons}}  ES \nonumber \\
ES  \overset{k_2}{\rightarrow}   E+ P  \label{eq8} \\
ES  +   S \underset{k_{-3}}{\stackrel{k_3}{\rightleftharpoons}}  ESS
\nonumber
\end{eqnarray}
there are five reactive processes (five channels in the language of the Gillespie algorithm) and consequently five families of counting processes $\{\chi_{i_h}^{(h)}(t,\cdot) \}$, $h=1,\dots,5$, should
be introduced, so that the formulation of the discrete stochastic dynamics reads
\begin{eqnarray}
\frac{d N_S(t)}{d t} & = & - \sum_{i=1}^{N_S(t)} \frac{d \chi_i^{(1)}(t, \widetilde{k}_1 \, N_E(t))}{d t} +
\sum_{j=1}^{N_{ES}(t)} \frac{d \chi_j^{(2)}(t, k_{-1})}{d t}  \nonumber \\
\frac{d N_E(t)}{d t} & = & - \sum_{i=1}^{N_S(t)} \frac{d \chi_i^{(1)}(t, \widetilde{k}_1 \, N_E(t))}{d t} +
\sum_{j=1}^{N_{ES}(t)} \frac{d \chi_j^{(2)}(t, k_{-1})}{d t}   + \sum_{h=1}^{N_{ES}(t)} \frac{d \chi_h^{(3)}(t, k_2)}{d t} \nonumber \\
\frac{d N_{ES}(t)}{d t} & = &  \sum_{i=1}^{N_S(t)} \frac{d \chi_i^{(1)}(t, \widetilde{k}_1 \, N_E(t))}{d t} -
\sum_{j=1}^{N_{ES}(t)} \frac{d \chi_j^{(2)}(t, k_{-1})}{d t}   - \sum_{h=1}^{N_{ES}(t)} \frac{d \chi_h^{(3)}(t, k_2)}{d t}
 -   \sum_{k=1}^{N_S(t)} \frac{d \chi_k^{(4)}(t, \widetilde{k}_3 \, N_{ES}(t))}{d t} \nonumber \\
&+ & \sum_{l=1}^{N_{ESS}(t)} \frac{d \chi_l^{(5)}(t, k_{-3})}{d t}  \label{eq9} \\
\frac{d N_{ESS}(t)}{d t} & = & \sum_{k=1}^{N_S(t)} \frac{d \chi_k^{(4)}(t, \widetilde{k}_3 \, N_{ES}(t))}{d t}
-  \sum_{l=1}^{N_{ESS}(t)} \frac{d \chi_l^{(5)}(t, k_{-3})}{d t} \nonumber \\
\frac{d N_P(t)}{d t} & = &  \sum_{h=1}^{N_{ES}(t)} \frac{d \chi_h^{(3)}(t, k_2)}{d t}
\nonumber
\end{eqnarray}
equipped with the initial conditions $c_S(0)=c_{S,0}$, $c_{E}(0)=c_{E,0}$, $c_{ES}(0)=c_{ESS}(0)=c_P(0)=0$. Observe that  for the bimolecular steps we have
used a number-dependent rate coefficient. This is just one  possibility, out 
of other fully equivalent alternatives, of defining bimolecular
reacting processes, and out of tem a numerical algorithm for solving them. 
This issue, and its computational implications will
be addressed elsewhere \cite{PG_next}.
The granularity number $N_g$ can be fixed, so that
\begin{equation}
N_S(0)= [ c_{S,0} \, N_g ] \, , \qquad N_{E,0}=[c_{E,0} \, N_g]
\label{eq10}
\end{equation}
where $[\xi ]$ indicates the integer part of $\xi$,
thus defining the relation betwen $N_\alpha(t)$ and $c_\alpha(t)$, namely
$c_\alpha(t)= N_\alpha(t)/N_g$, $\alpha=S,\, E,\, ES,\,ESS,\,P$. This implies also that the effective rate parameters
entering the discrete stochastic evolution equation (\ref{eq9}), and 
associated with the
two bimolecular reactive steps, are given by $\widetilde{k}_1=k_1/N_g$, and $\widetilde{k}_3=k_3/N_g$.

\begin{figure}
\includegraphics[width=10cm]{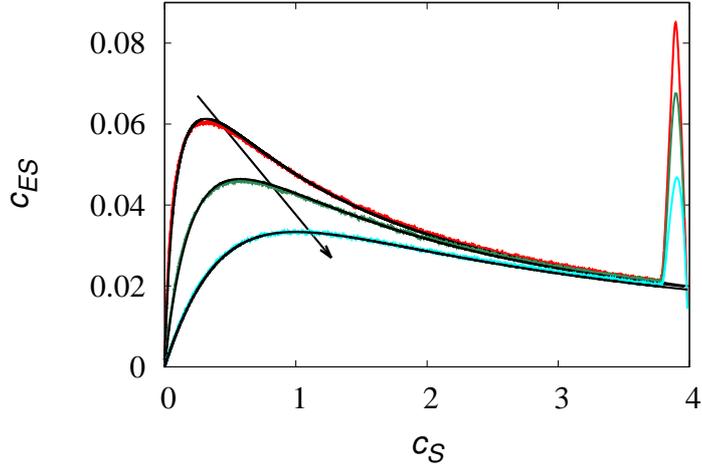}
\caption{$c_{ES}$ vs $c_S$ plot of the substrate-inhibited  enzymatic kinetics discussed in the main text. Symbols (in color) are
the results of stochastic simulations of a single realization of the process eq. (\ref{eq8}), (black) solid lines
the graph of the quasi steady-state approximation. The arrow indicates increasing values of $K_M$, i.e. decreasing values of
$k_1=20,\, 6,\, 2$.}
\label{Fig3}
\end{figure}

Consider the case $k_{-1}=k_2=k_3=k_{-3}=1$, $c_{S,0}=4$, $c_{E,0}=0.1$. In this case
the quasi steady-state approximation of the $c_{ES}$-$c_S$ curve (representing the slow manifold of the kinetics
takes the expression
\begin{equation}
c_{ES}= \frac{ c_{E,0} \, c_S}{K_M+ c_S + \beta \, c_S^2} \,, \qquad K_M= \frac{ k_{-1}+k_2}{k_1} \, , \quad
\beta= \frac{k_{-3}}{k_3}
\label{eq11}
\end{equation}
Figure  \ref{Fig3} depicts the $c_{ES}$-$c_S$ graph obtained from a single realization
of the stochastic process eq. (\ref{eq8}) at several values of $k_1$ so as to modify the Michaelis-Menten constant $K_M$ for a value $N_g=10^6$
of the granularity number.

Apart from the initial transient giving rise to an overshot in the values of $c_{ES}$ near $c_S \simeq c_{S,0}$, the dynamics rapidly
collapses towards the slow manifold and the stochastic simulations at high $N_g$-value provide a reliable description
of the mean-field behavior starting from a single stochastic realization. 

To conclude, we want to point out some advantages and extensions of the present
approach:
\begin{itemize}
\item it shows a direct analogy  between chemical reaction kinetics,
radiative processes and stochastic formulation of open quantum systems, thus, paving the way for a unified treatment of the interpaly between these phenomena,
that is particularly important in the field of photochemistry, and in the
foundation of statistical physics \cite{PGradia,petruccione};
\item it can be easily extended to semi-Markov transition. This is indeed
the case of the growth kinetics of eukaryotic microorganisms, the physiological state  of which
can be parametrized with respect to internal (hidden) parameters such as the
age, the cytoplasmatic content, etc.;
\item it can be easily extended to include transport phenomena. In point 
of fact, the occurrence of Markovian or semi-Markovian transitions
in modeling chemical kinetics is analogous to the
transitions occurring in the direction of motion (Poisson-Kac processes,
L\'evy flights, Extended Poisson-Kac processes) or in the 
velocity (linearized Boltzmannian schemes) \cite{GPK,EPK,Levy}.
\item it is closely related
to the formulation of  stochastic differential  equations for the
thermalization of athermal system \cite{athermo1}, in which
the classical mesoscopic  description of thermal  fluctuations, using
the increments of a Wiener process, is replaced by a dynamic model
involving the increments of a counting process.
\end{itemize}
Due to the limitations of a Letter, all these issues will be addressed in forthcoming works.
But apart for these extensions and improvements, the proposed formulation indicates that the stochastic theory of chemical reactions can be built upon
a simple and consistent mathematical formalism describing the elementary
reactive events as  Markovian or semi-Markovian counting processes
\cite{semimarkovcounting}, that perfectly fits with the description of 
molecular 
non reactive events (molecular collisions), providing an unifying stochastic
formalism of elementary (classical and quantum) molecular events.

\end{document}